\documentstyle[11pt,moriond,epsfig]{article}

\def\be{\begin{equation}}
\def\ee{\end{equation}}

\def\gsim{\mathrel{%
\rlap{\raise 0.511ex \hbox{$>$}}{\lower 0.511ex
\hbox{$\sim$}}}}

\def\lsim{\mathrel{
\rlap{\raise 0.511ex \hbox{$<$}}{\lower 0.511ex
\hbox{$\sim$}}}}

\begin{document}

\vspace*{4cm}

\title{EQUIVALENCE PRINCIPLE AND CLOCKS}

\vglue 1.5cm

\author{T. DAMOUR}

\address{Institut des Hautes Etudes Scientifiques, 91440 
Bures-sur-Yvette, France\\
and DARC, CNRS - Observatoire de Paris, 92195
Meudon Cedex, France}

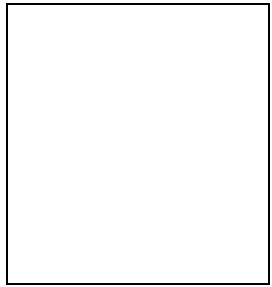
\maketitle\abstracts{
String theory suggests the existence of gravitational-strength
scalar fields (``dilaton'' and ``moduli'') whose couplings to
matter violate the equivalence principle. This provides a new
motivation for high-precision clock experiments, as well as a
generic theoretical framework for analyzing their significance.
}

\section{Introduction}

The equivalence principle was postulated by Einstein as a 
foundation stone for general relativity. The equivalence 
principle stipulates that the long-range gravitational 
interaction is entirely described by a universal coupling of 
``matter'' (leptons, quarks, gauge fields and Higgs fields) to a 
(dynamical) second-rank symmetric tensor field $g_{\mu \nu} 
(x^{\lambda})$, replacing everywhere in the matter Lagrangian the 
usual, kinematical, special relativistic (Minkowski) metric 
$\eta_{\mu \nu}$. This principle assumes that all the 
non-gravitational (dimensionless) coupling constants of matter 
(gauge couplings, CKM mixing angles, mass ratios,$\ldots$) are 
non-dynamical, i.e. take (at least at large distances) some fixed 
(vacuum expectation) values, independently of where and when, in 
spacetime, they are measured. Two of the best experimental tests 
of the equivalence principle are:

(i) tests of the universality of free fall, i.e. the fact that 
all bodies fall with the same acceleration in an external 
gravitational field; and

(ii) tests of the ``constancy of the constants''.

Laboratory experiments (due notably, in our century, to 
E\"otv\"os, Dicke, Braginsky and Adelberger) have verified the 
universality of free fall to the $10^{-12}$ level. For instance, 
the fractional difference in free fall acceleration of Beryllium 
and Copper samples was found to be~\cite{Su94}
\be
\left( \frac{\Delta a}{a} \right)_{\rm Be \, Cu} = (-1.9 \pm
2.5) \times 10^{-12} \, . \label{eq:adel}
\ee

The Lunar Laser Ranging experiment~\cite{LLR} has also verified that the Moon 
and the Earth fall with the same acceleration toward the Sun to better than one 
part in $10^{12}$
\be
\left( \frac{\Delta a}{a} \right)_{\rm Moon \, Earth} = (-3.2 \pm
4.6) \times 10^{-13} \, . \label{eq:llr}
\ee

On the other hand, a recent reanalysis of the Oklo phenomenon (a 
natural fission reactor which operated two billion years ago in 
Gabon, Africa) gave a very tight limit on a possible time 
variation of the fine-structure ``constant'', namely~\cite{DD96}
\be
-0.9 \times 10^{-7} < \frac{e_{\rm Oklo}^2 - e_{\rm now}^2}{e^2} 
< 1.2 \times 10^{-7} \, , \label{eq:oklo1}
\ee
\be
-6.7 \times 10^{-17} \, {\rm yr}^{-1} < \frac{d}{dt} \ {\rm
ln} \ e^2 < 5.0 \times 10^{-17} \, {\rm yr}^{-1} \, .
\label{eq:oklo2}
\ee
The tightness of the experimental limits (\ref{eq:adel})--(\ref{eq:oklo2}) might 
suggest 
to apply Occam's razor and to declare that the equivalence 
principle must be exactly enforced. However, the theoretical 
framework of modern unification theories, and notably string 
theory, suggest that the equivalence principle must be violated. 
Even more, the type of violation of the equivalence principle 
suggested by string theory is deeply woven into the basic fabric 
of this theory. Indeed, string theory is a very ambitious attempt 
at unifying all interactions within a consistent quantum 
framework. A deep consequence of string theory is that 
gravitational and gauge couplings are unified. In intuitive 
terms, while Einstein proposed a framework where geometry and 
gravitation were becoming united as a dynamical field $g_{\mu \nu} 
(x)$, i.e. a soft structure influenced by the presence of matter, 
string theory extends this idea by proposing a framework where 
geometry, gravitation, gauge couplings, and gravitational 
couplings all become soft structures described by interrelated 
dynamical fields. A symbolic equation expressing this softened, 
unified structure is
\be
g_{\mu \nu} (x) \sim g^2 (x) \sim G(x) \, . \label{eq:ggg}
\ee
It is conceptually pleasing to note that string theory proposes 
to render dynamical the structures left rigid (or 
kinematical) by general relativity. Technically, 
Eq.~(\ref{eq:ggg}) refers to the fact that string theory (as well 
as Kaluza-Klein theories) predicts the existence, at a 
fundamental level, of scalar partners of Einstein's tensor field 
$g_{\mu \nu}$, the model-independent ``dilaton'' field $\Phi 
(x)$, and various ``moduli fields''. The dilaton field, notably, 
plays a crucial role in string theory in that it determines the 
basic ``string coupling constant'' $g_s = e^{\Phi (x)}$, which 
determines in turn the (unified) gauge and gravitational coupling 
constants $g \sim g_s$, $G \propto g_s^2$, as exemplified by the 
low-energy effective action
\be
L_{\rm eff} = e^{-2\Phi} \left[ \frac{R(g)}{\alpha'} + 
\frac{4}{\alpha'} \, (\nabla \Phi)^2 - \frac{1}{4} \, F_{\mu 
\nu}^2 - i \overline{\psi} \, D \psi - \ldots \right] \, . 
\label{eq:eff}
\ee

A softened structure of the type of Eq.~(\ref{eq:ggg}), embodied 
in the effective action (\ref{eq:eff}), implies a deep violation 
of Einstein's equivalence principle. Bodies of different nuclear 
compositions fall with different accelerations because, for 
instance, the part of the mass of nucleus $A$ linked to the 
Coulomb interaction of the protons depends on the space-variable 
fine-structure constant $e^2 (x)$ in a non-universal, 
composition-dependent manner. This raises the problem of the 
compatibility of the generic string prediction (\ref{eq:ggg}) 
with experimental tests of the equivalence principle, such as 
Eqs.~(\ref{eq:adel}), (\ref{eq:llr}) or (\ref{eq:oklo2}). It is 
often assumed that the softness (\ref{eq:ggg}) applies only at 
short distances, because the dilaton and moduli fields are likely 
to acquire a non zero mass after supersymmetry breaking. However, 
a mechanism has been proposed~\cite{DP} to reconcile
in a natural manner the existence of a {\it massless} dilaton (or 
moduli) field as a
fundamental partner of the graviton field $g_{\mu \nu}$ with
the current level of precision $(\sim 10^{-12})$ of
experimental tests of the equivalence principle. In the
mechanism of~\cite{DP} (see also~\cite{DN} for
metrically-coupled scalars) the very small couplings necessary
to ensure a near universality of free fall, $\Delta a/a <
10^{-12}$, are dynamically generated by the expansion of the
universe, and are compatible with couplings ``of order unity''
at a fundamental level.

The aim of the present paper is to emphasize the rich
phenomenological consequences of long-range dilaton-like fields, and the fact 
that high-precision clock experiments might contribute to searching for, or 
constraining, their existence.
More precisely, the basic question we wish to address here is the 
following: given
the existing experimental tests of gravity, and given the
currently favored theoretical framework, can high-precision
clock experiments probe interesting theoretical possibilities
which remain yet unconstrained ? In addressing this question
we wish to assume, as theoretical framework, the class of
effective field theories suggested by string theory.

For historical completeness, let us mention that the theoretical 
framework which has been most considered in the phenomenology of 
gravitation, i.e. the class of
``metric'' theories of gravity~\cite{W81}, which includes most
notably the ``Brans-Dicke''-type tensor-scalar theories,
appears, from a modern perspective, as being rather artificial. 
This is good news because
the phenomenology of ``non metric'' theories is richer and
offers new possibilities for clock experiments. Historically,
the restricted class of ``metric'' theories was introduced in
1956 by Fierz~\cite{F56} to prevent, in an {\it ad hoc} way,
too violent a conflict between experimental tests of the
equivalence principle and the existence of a scalar
contribution to gravity as suggested by the theories of
Kaluza-Klein~\cite{KK} and Jordan~\cite{J}. Indeed, Fierz was
the first one to notice that a Kaluza-Klein scalar would
generically strongly violate the equivalence principle. He
then proposed to restrict artificially the couplings of the
scalar field to matter so as to satisfy the equivalence
principle. The restricted class of
equivalence-principle-preserving couplings introduced by Fierz
is now called ``metric'' couplings. Under the aegis of Dicke,
Nordtvedt, Thorne and Will a lot of attention has been given
to ``metric'' theories of gravity\footnote{Note, however, that
Nordtvedt, Will, Haugan and others (for references see~\cite{W81}) studied 
conceivable phenomenological
consequences of generic ``non metric'' couplings, without
using a motivated field-theory
framework to describe such couplings.}, and notably to their
quasi-stationary-weak-field phenomenology (``PPN framework'',
see, e.g.,~\cite{W81}).

\section{Generic effective theory of a long-range dilaton}

Motivated by string theory, we consider the generic class of
theories containing a long-range dilaton-like scalar field
$\varphi$. The effective Lagrangian describing these theories
has the form (after a conformal transformation to the ``Einstein 
frame''):
\begin{eqnarray}
L_{\rm eff} &=& \frac{1}{4q} R(g_{\mu\nu}) - \frac{1}{2q} \
(\nabla \varphi)^2 - \frac{1}{4e^2 (\varphi)} \ (\nabla_{\mu}
A_{\nu} - \nabla_{\nu} A_{\mu})^2 \nonumber \\
&-& \sum_A \ \left[\overline{\psi}_A \,
\gamma^{\mu} (\nabla_{\mu} -iA_{\mu}) \psi_A + m_A (\varphi)
\, \overline{\psi}_A \psi_A \right] + \cdots \label{eq:01}
\end{eqnarray}
Here, $q\equiv 4\pi \, \overline G$ where $\overline G$ denotes
a bare Newton's constant, $A_{\mu}$ is the electromagnetic
field, and $\psi_A$ a Dirac field describing some fermionic
matter. At the low-energy, effective level (after the breaking
of $SU(2)$ and the confinement of colour), the coupling of the
dilaton $\varphi$ to matter is described by the
$\varphi$-dependence of the fine-structure ``constant'' $e^2
(\varphi)$ and of the various masses $m_A (\varphi)$. Here,
$A$ is a label to distinguish various particles. [A deeper
description would include more coupling functions, e.g.
describing the $\varphi$-dependences of the $U(1)_Y$,
$SU(2)_L$ and $SU(3)_c$ gauge coupling ``constants''.]

The strength of the coupling of the dilaton $\varphi$ to the
mass $m_A (\varphi)$ is given by the quantity
\be
\alpha_A \equiv \frac{\partial \ {\rm ln} \ m_A
(\varphi_0)}{\partial \ \varphi_0} \, , \label{eq:02}
\ee
where $\varphi_0$ denotes the ambient value of $\varphi (x)$
(vacuum expectation value of $\varphi (x)$ around the mass
$m_A$, as generated by external masses and cosmological
history). For instance, the usual PPN parameter $\gamma -1$
measuring the existence of a (scalar) deviation from the pure
tensor interaction of general relativity is given 
by~\cite{DEF},~\cite{DP}
\be
\gamma -1 = -2 \ \frac{\alpha_{\rm had}^2}{1+\alpha_{\rm had}^2}
\, , \label{eq:03}
\ee
where $\alpha_{\rm had}$ is the (approximately universal)
coupling (\ref{eq:02}) when $A$ denotes any (mainly) hadronic
object.

The Lagrangian (\ref{eq:01}) also predicts (as discussed 
in~\cite{DP}) a link between the coupling strength (\ref{eq:02})
and the violation of the universality of free fall:
\be
\frac{a_A -a_B}{\frac{1}{2} (a_A + a_B)} \simeq (\alpha_A
-\alpha_B) \alpha_E \sim -5\times 10^{-5} \, \alpha_{\rm
had}^2 \, . \label{eq:04}
\ee
Here, $A$ and $B$ denote two masses falling toward an external
mass $E$ (e.g. the Earth), and the numerical factor $-5 \times
10^{-5}$ corresponds to $A= {\rm Be}$ and $B= {\rm Cu}$. The
experimental limit Eq.~(\ref{eq:adel})
shows that the (mean hadronic) dilaton coupling strength is
already known to be very small:
\be
\alpha_{\rm had}^2 \lsim 10^{-7} \, . \label{eq:06}
\ee

Free fall experiments, such as Eq.~(\ref{eq:adel}) or the
comparable Lunar Laser Ranging constraint Eq.~(\ref{eq:llr}), 
give the
tightest constraints on any long-range dilaton-like coupling.
Let us mention, for comparison, that solar-system measurements
of the PPN parameters (as well as binary pulsar measurements)
constrain the dilaton-hadron coupling to $\alpha_{\rm had}^2 <
10^{-3}$ (recently announced VLBI measurements improve this 
constraint to the $2 \times 10^{-4}$ level), while the best 
current constraint on the time
variation of the fine-structure ``constant'' (deduced from the
Oklo phenomenon), namely Eq.~(\ref{eq:oklo2}),
yields from Eq. (\ref{eq:16}) below, $\alpha_{\rm had}^2 \lsim
3 \times 10^{-4}$. [For an updated review of experimental tests 
of gravity, see the chapter 14 of the Review of Particle Physics, 
available on http://pdg.lbl.gov/]

To discuss the probing power of clock experiments, we need
also to introduce other coupling strengths, such as
\be
\alpha_{\rm EM} \equiv \frac{\partial \ {\rm ln} \ e^2
(\varphi_0)}{\partial \ \varphi_0} \, , \label{eq:07}
\ee
measuring the $\varphi$-variation of the electromagnetic (EM)
coupling constant\footnote{Note that we do not use the
traditional notation $\alpha$ for the fine-structure constant
$e^2 / 4\pi \hbar c$. We reserve the letter $\alpha$ for
denoting various dilaton-matter coupling strengths. Actually,
the latter coupling strengths are analogue to $e$ (rather than
to $e^2$), as witnessed by the fact that observable deviations
from Einsteinian predictions are proportional to products of
$\alpha$'s, such as $\alpha_A \alpha_E$, $\alpha_{\rm had}^2$,
etc$\ldots$}, and
\be
\alpha_A^{A^*} \equiv \frac{\partial \ {\rm ln} \ E_A^{A^*}
(\varphi_0)}{\partial \ \varphi_0} \, , \label{eq:08}
\ee
where $E_A^{A^*}$ is the energy difference between two atomic
energy levels.

In principle, the quantity $\alpha_A^{A^*}$ can be expressed
in terms of more fundamental quantities such as the ones
defined in Eqs. (\ref{eq:02}) and (\ref{eq:07}). For instance,
in an hyperfine transition
\be
E_A^{A^*} \propto (m_e \, e^4) \ g_I \ \frac{m_e}{m_p} \ e^4 \
F_{\rm rel} (Z e^2) \, , \label{eq:09}
\ee
so that
\be
\alpha_A^{A^*} \simeq 2 \, \alpha_e -\alpha_p + \alpha_{\rm EM}
\left( 4+\frac{d \ {\rm ln} \ F_{\rm rel}}{d \ {\rm ln} \ e^2}
\right) \, . \label{eq:10}
\ee
Here, the term $F_{\rm rel} (Z e^2)$ denotes the relativistic
(Casimir) correction factor~\cite{Casimir}. Moreover, in any
theory incorporating gauge unification one expects to have the
approximate link~\cite{DP}
\be
\alpha_A \simeq \left( 40.75 - {\rm ln} \ \frac{m_A}{1 \ {\rm
GeV}} \right) \ \alpha_{\rm EM} \, , \label{eq:11}
\ee
at least if $m_A$ is mainly hadronic.

\section{Clock experiments and dilaton couplings}

The coupling parameters introduced above allow one to describe
the deviations from general relativistic predictions in most
clock experiments~\cite{TD}. Let us only mention some simple
cases.

First, it is useful to distinguish between ``global'' clock
experiments where one compares spatially distant clocks, and
``local'' clock experiments where the clocks being compared
are next to each other. The simplest global clock experiment
is a static redshift experiment comparing (after transfer by
electromagnetic links) the frequencies of the same transition
$A^* \rightarrow A$ generated in two different locations ${\bf
r}_1$ and ${\bf r}_2$. The theory of Section 2 predicts a
redshift of the form (we use units in which $c=1$)
\be
\frac{\nu_A^{A^*} ({\bf r}_1)}{\nu_A^{A^*} ({\bf r}_2)} \simeq
1 + (1 + \alpha_A^{A^*} \, \alpha_E) \ (\overline{U}_E ({\bf
r}_2) - \overline{U}_E ({\bf r}_1)) \, , \label{eq:12}
\ee
where
\be
\overline{U}_E \, ({\bf r}) = \frac{\overline G \, m_E}{r}
\label{eq:13}
\ee
is the {\it bare} Newtonian potential generated by the
external mass $m_E$ (say, the Earth). Such a result has the
theoretical disadvantage of depending on other experiments for
its interpretation. Indeed, the {\it bare} potential
$\overline{U}_E$ is not directly measurable. The measurement
of the Earth potential by the motion of a certain mass $m_B$
gives access to $(1+\alpha_B \, \alpha_E) \ \overline{U}_E \,
({\bf r})$. The theoretical significance of a global clock
experiment such as (\ref{eq:12}) is therefore fairly indirect,
and involves other experiments and other dilaton couplings. One
can generalize (\ref{eq:12}) to a more general, non static
experiment in which different clocks in relative motion are
compared. Many different ``gravitational potentials'' will
enter the result, making the theoretical significance even
more involved.

A conceptually simpler (and, probably, technologically less
demanding) type of experiment is a differential, ``local''
clock experiment. Such ``null'' clock experiments have been
proposed by Will~\cite{W81} and first performed by Turneaure
et al.~\cite{T83}. The theoretical significance of such
experiments within the context of dilaton theories is much
simpler than that of global experiments. For instance if
(following the suggestion of~\cite{PTM}) one locally compares
two clocks based on hyperfine transitions in alkali atoms with
different atomic number $Z$, one expects to find a ratio of
frequencies 
\be
\frac{\nu_A^{A^*} ({\bf r})}{\nu_B^{B^*} ({\bf r})} \simeq
\frac{F_{\rm rel} (Z_A \, e^2 (\varphi_{\rm loc}))}{F_{\rm
rel} (Z_B \, e^2 (\varphi_{\rm loc}))} \, , \label{eq:14}
\ee
where the local, ambient value of the dilaton field
$\varphi_{\rm loc}$ might vary because of the (relative)
motion of external masses with respect to the clocks
(including the effect of the cosmological expansion). The
directly observable fractional variation of the ratio
(\ref{eq:14}) will consist of two factors:
\be
\delta \ {\rm ln} \ \frac{\nu_A^{A^*}}{\nu_B^{B^*}} = \left[
\frac{\partial \ {\rm ln} \ F_{\rm rel} (Z_A \, e^2)}{\partial
\ {\rm ln} \ e^2} - \frac{\partial \ {\rm ln} \ F_{\rm rel}
(Z_B \, e^2)}{\partial \ {\rm ln} \ e^2} \right] \times \delta
\ {\rm ln} \ e^2 \, . \label{eq:15}
\ee
The ``sensitivity'' factor in brackets due to the
$Z$-dependence of the Casimir term can be made of order 
unity~\cite{PTM}, while the fractional variation of the
fine-structure constant is expected in dilaton theories to be
of order~\cite{DP},~\cite{TD}
\begin{eqnarray}
\delta \ {\rm ln} \ e^2 (t) &=& -2.5 \times 10^{-2} \
\alpha_{\rm had}^2 \ U(t) \nonumber \\
&-& 4.7 \times 10^{-3} \ \kappa^{-1/2} ({\rm tan} \ \theta_0)
\ \alpha_{\rm had}^2 \ H_0 (t-t_0) \, . \label{eq:16}
\end{eqnarray}
Here, $U(t)$ is the value of the externally generated
gravitational potential at the location of the clocks, and
$H_0 \simeq 0.5 \times 10^{-10} \ {\rm yr}^{-1}$ is the Hubble
rate of expansion. [The factor $\kappa^{-1/2} \ {\rm tan} \
\theta_0$ is expected to be $\sim 1$.]

The (rough) theoretical prediction (\ref{eq:16}) allows one to
compare quantitatively the probing power of clock experiments
to that of equivalence principle tests. Let us
(optimistically) assume that clock stabilities of order
$\delta \nu / \nu \sim 10^{-17}$ (for the relevant time scale)
can be achieved. A differential {\it ground} experiment (using
the variation of the Sun's potential due to the Earth
eccentricity) would probe the level $\alpha_{\rm had}^2 \sim
3\times 10^{-6}$. A geocentric satellite differential
experiment could probe $\alpha_{\rm had}^2 \sim 5\times
10^{-7}$. These levels are impressive (compared to present
solar-system tests of the PPN parameter $\gamma$ giving the
constraint $\alpha_{\rm had}^2 \simeq (1-\gamma) / 2 <
2 \times 10^{-4}$), but are not as good as the present
equivalence-principle limit (\ref{eq:06}). To beat the level
(\ref{eq:06}) one needs to envisage an heliocentric
differential clock experiment (a few solar radii probe within
which two hyper-stable clocks are compared). Such an
experiment could, according to Eq. (\ref{eq:16}), reach the
level $\alpha_{\rm had}^2 \sim 10^{-9}$. [Let us also note
that a gravitational time delay global experiment using clocks
beyond the Sun as proposed by C. Veillet (SORT concept) might
(optimistically) probe the level $\alpha_{\rm had}^2 \sim
10^{-7}$.] It is, however, to be noted that a much refined test
of the equivalence principle such as STEP (Satellite Test of
the Equivalence Principle) aims at measuring $\Delta a/a \sim
10^{-18}$ which corresponds to the level $\alpha_{\rm had}^2
\sim 10^{-14}$, i.e. five orders of magnitude better than any
conceivable clock experiment.

\section{Conclusions}

In summary, the main points of the present contribution are:
\begin{enumerate}
\item[$\bullet$] Independently of any theory, the result
(\ref{eq:oklo2}) of a recent reanalysis of the Oklo 
phenomenon~\cite{DD96} gives a motivation, and a target, for 
improving
laboratory clock tests of the time variation of the
fine-structure constant $e^2$ (which are at the $3.7 \times
10^{-14} \ {\rm yr}^{-1}$ level~\cite{PTM}).
\item[$\bullet$] Modern unification theories, and especially
string theory, suggest the existence of new
gravitational-strength fields, notably scalar ones
(``dilaton'' or ``moduli''), whose couplings to matter violate
the equivalence principle. These fields would induce a
spacetime variability of the coupling constants of physics
(such as the fine-structure constant). High-precision clock
experiments are excellent probes of such a possibility.
\item[$\bullet$] The generic class of dilaton theories defined
in Section 2 provides a well-defined theoretical framework in
which one can discuss the phenomenological consequences of the
existence of a dilaton-like field. Such a theoretical
framework (together with some assumptions, e.g. about gauge
unification and the origin of mass hierarchy) allows one to
compare and contrast the probing power of clock experiments to
that of other experiments.
\item[$\bullet$] Local, differential clock experiments (of the
``null'' type of~\cite{T83}) appear as conceptually cleaner,
and technologically less demanding, probes of
dilaton-motivated violations of the equivalence principle than
global, absolute clock experiments (of the Gravity Probe A
type).
\item[$\bullet$] If we use the theoretical assumptions of
Section 2 to compare clock experiments to free-fall
experiments, one finds that one needs to send and intercompare
two ultra-high-stability clocks in near-solar orbit in order
to probe dilaton-like theories more deeply than {\it present}
free-fall experiments. Currently proposed improved satellite
tests of the equivalence principle would, however, beat any
clock experiment in probing even more deeply such theories.
\item[$\bullet$] At the qualitative level, it is, however,
important to note that clock experiments (especially of the
``global'', GPA type) probe different combinations of basic
coupling parameters than free-fall experiments. This is
visible in Eq. (\ref{eq:10}) which shows that $\alpha_A^{A^*}$
contains the leptonic quantity $\alpha_e = \partial \ {\rm ln}
\ m_{\rm electron} / \partial \ \varphi_0$ without any small
factor\footnote{Free-fall experiments couple predominantly to
hadronic quantities such as $\alpha_p =
\partial \ {\rm ln} \ m_{\rm proton} / \partial \ \varphi_0$,
and to Coulomb-energy effects proportional to $\alpha_{\rm
EM}$. The effect of the leptonic quantity $\alpha_e$ is down
by a small factor $\sim m_e / m_p \sim 1/1836$.}.
\end{enumerate}


\section*{References}
\begin{thebibliography}{9999999}
\bibitem{Su94} Y. Su et al., {\it Phys. Rev.} D {\bf 50}, 3614 
(1994).

\bibitem{LLR} J.O. Dickey et al., {\it Science} {\bf 265}, 482 
(1994); 

J.G. Williams et al., {\it Phys. Rev.} D {\bf 53}, 6730 (1996).

\bibitem{DD96} T. Damour and F. Dyson, {\it Nucl.
Phys.} B {\bf 480}, 37 (1996); hep-ph/9606486.

\bibitem{DP} T. Damour and A.M. Polyakov, {\it Nucl. Phys.} B {\bf
423}, 532 (1994); {\it Gen. Rel. Grav.} {\bf 26}, 1171 (1994).

\bibitem{DN} T. Damour and K. Nordtvedt, {\it Phys. Rev. Lett.} 
{\bf 70}, 2217 (1993); {\it Phys. Rev.} D {\bf 48}, 3436 (1993).

\bibitem{W81} C.M. Will, {\it Theory and experiment in
gravitational physics} (Cambridge University Press, 1981);
second edition 1993.

\bibitem{F56} M. Fierz, {\it Helv. Phys. Acta} {\bf 29}, 128 
(1956).

\bibitem{KK} T. Kaluza, {\it Sitz. der K. Preuss. Akad. der 
Wiss.}, 966 (1921);

O. Klein, {\it Zeit. f\"ur Physik} {\bf 37}, 895 (1926).

\bibitem{J} P. Jordan, {\it Nature} {\bf 164}, 637 (1949); {\it
Schwerkraft und Weltall} (Vieweg, Braunschweig, 1955).

\bibitem{DEF} T. Damour and G. Esposito-Far\`ese, {\it Class.
Quant. Grav.} {\bf 9}, 2093 (1992).

\bibitem{Casimir} H. Kopfermann, {\it Nuclear Moments}
(Academic Press, New York, 1958), pp.123-138;

H.B.G. Casimir, {\it On the Interaction between Atomic Nuclei
and Electrons} (Freeman, San Francisco, 1963), p.54.

\bibitem{TD} T. Damour, in preparation.

\bibitem{T83} J.P. Turneaure et al., {\it Phys. Rev.} D {\bf 27},
1705 (1983).

\bibitem{PTM} J.D. Prestage, R.L. Tjoelker and L. Maleki,
{\it Phys. Rev. Lett.} {\bf 74}, 3511 (1995).

\end{thebibliography}
\end{document}